\documentclass[prb,aps,showpacs,twocolumn,preprintnumbers,amsmath,amssymb,superscriptaddress]{revtex4-1}
\usepackage{amsmath,amssymb,amsfonts}
\usepackage{graphicx}
\usepackage[colorlinks=True,linkcolor=red,citecolor=blue,urlcolor=blue]{hyperref}
\usepackage{xcolor}
\usepackage{chngcntr}
\usepackage{braket}
\usepackage{hyperref}
\usepackage{nicefrac}
\usepackage{bm}
\usepackage{multirow}
\usepackage{ulem}
\usepackage{physics}
\usepackage{amsmath}
\usepackage{tikz}
\usepackage{mathdots}
\usepackage{yhmath}
\usepackage{cancel}
\usepackage{color}
\usepackage{siunitx}
\usepackage{array}
\usepackage{multirow}
\usepackage{amssymb}
\usepackage{gensymb}
\usepackage{tabularx}
\usepackage{extarrows}
\usepackage{listings}
\usepackage{booktabs}
\usetikzlibrary{fadings}
\usetikzlibrary{patterns}
\usetikzlibrary{shadows.blur}
\usetikzlibrary{shapes}

\begin{document}
\bibliographystyle{apsrev4-1}
\title{Charge transfer in T/H heterostructures of transition metal dichalcogenides}

\author{Irián Sánchez-Ramírez}
\affiliation{Donostia International Physics Center, P. Manuel de Lardizabal 4, 20018 Donostia-San Sebastian, Spain}
\affiliation{Departamento de Fisica de Materiales, Facultad de Ciencias Quimicas, Universidad del Pais Vasco (UPV-EHU), P. Manuel de Lardizabal 3, 20018 Donostia-San Sebastian, Spain}
\author{Maia G. Vergniory}
\affiliation{Donostia International Physics Center, P. Manuel de Lardizabal 4, 20018 Donostia-San Sebastian, Spain}
\affiliation{Max Planck Institute for Chemical Physics of Solids, 01187 Dresden, Germany}
\author{Fernando de Juan}
\affiliation{Donostia International Physics Center, P. Manuel de Lardizabal 4, 20018 Donostia-San Sebastian, Spain}
\affiliation{Departamento de Fisica de Materiales, Facultad de Ciencias Quimicas, Universidad del Pais Vasco (UPV-EHU), P. Manuel de Lardizabal 3, 20018 Donostia-San Sebastian, Spain}
\affiliation{IKERBASQUE, Basque Foundation for Science, Maria Diaz de Haro 3, 48013 Bilbao, Spain}

\date{\today}
%%%%%%%%%%%%%%%%%%%%%%%%%%%%%%%%%%%%%%%%%%%%%%%%%%%%%%%%%%%%%%%%%%%%%%%%%%%%%
\begin{abstract}
The $\sqrt{13}\times\sqrt{13}$ charge density wave state of the T polytype of MX$_2$ (M=Nb,Ta, X=S, Se) is known to host a half-filled flat band, which electronic correlations drive into a Mott insulating state. When T polytypes are coupled to strongly metallic H polytypes, such as in T/H bilayer heterostructures or the bulk 4H$_b$ polytype, charge transfer can destabilize the Mott state, but quantifying its magnitude has been a source of controversy. In this work, we perform a systematic ab-initio study of charge transfer for all experimentally relevant T/H bilayers and bulk 4H$_b$ structures. In all cases we find charge transfer from T to H layers which depends strongly on the interlayer distance but weakly on the Hubbard interaction. Additionally, Se compounds display smaller charge transfer than S compounds, and 4H$_b$ bulk polytypes display more charge transfer than isolated bilayers. We rationalize these findings in terms of band structure properties, and argue they might explain differences between compounds observed experimentally. Our work reveals the tendency to Mott insulation and the origin of superconductivity may vary significantly across the family of T/H heterostructures.
\end{abstract}
\maketitle

\section{Introduction}

Metallic transition metal dichalcogenides (TMD) MX$_2$ (M=Nb,Ta, X=S, Se)~\cite{Kalikhman73,Wilson75,Rossnagel11} are layered compounds made by stacking two basic MX$_2$ monolayer units known as H, with trigonal prismatic coordination of the metal, and T, with octahedral coordination~\footnote{As it is often done, we label the two-dimensional polytypes with letters T and H, and the bulk polytypes as 1T, 2H, 4H$_b$\ldots, where the number refers to the number of layers in the unit cell.}. Despite both having partially filled metal $d$-orbital bands, the low temperature properties of these building blocks are markedly different: while H layers show weak charge density waves (CDW) and remain metallic and often superconducting~\cite{Wilson75}, T layers are understood to be Mott insulators~\cite{Fazekas80}. The reason for this is the strong Star of David (SoD) $\sqrt{13} \times \sqrt{13}$ CDW reconstruction~\cite{Rossnagel06}, which gaps out most of the Fermi surface, leaving a  half-filled flat band at the Fermi level which derives from an isolated orbital located at the SoD centers. The presence of even moderate Coulomb repulsion in these orbitals should thus drive this system to a Mott insulator state~\cite{Fazekas80}, and perhaps a spin liquid~\cite{Law17}. The realization of such states in the bulk 1T polytype, made of stacked T layers, is however complicated by interlayer tunneling and the many possible CDW stacking patterns~\cite{Ritschel18}. However, the recent synthesis of T monolayers~\cite{Ruan21} has provided stronger support to the Mott insulator scenario. 

TMD heterostructures alternating T and H layers provide an interesting alternative to put the hypothesis of the Mott insulator to test. In these structures the SoD moments from the T layer are coupled to the metallic electrons from the H layer, providing a natural realization of a CDW-induced Kondo lattice. Recently synthesized T/H bilayers have indeed showed prominent zero bias peaks in TaSe$_2$~\cite{Ruan21,Wan22}, TaS$_2$~\cite{Vavno21,Ayani22} and NbSe$_2$~\cite{Liu21,Ganguli24} which have been interpreted as the Kondo effect. In addition to such isolated bilayers, two naturally occuring bulk polytypes formed by alternating T and H layers also exist, known as 4H$_b$~\cite{DiSalvo73,DiSalvo76} (with an inversion symmetric stacking of T/H bilayers) and 6R~\cite{Achari22,Pal23} (with an inversion breaking rhombohedral stacking). Interest in such compounds has been recently renewed because, in addition to displaying similar Kondo effects~\cite{Shen22,Nayak23}, 4H$_b$-TaS$_2$ shows unexpected signatures of unconventional superconductivity~\cite{Dentelski21}: spontaneous breaking of time-reversal symmetry at $T_c$ \cite{Ribak20}, spontaneous vortices in the superconducting state~\cite{Persky22}, superconducting edge modes \cite{Nayak21} and transport evidence of a two-component order parameter \cite{Almoalem22,Silber24}. 

The interpretation of all of these experiments and the relevance of the Mott insulator picture is critically dependent on two properties of the T/H interface which have remained under certain controversy: the interlayer hybridization $V$ and the charge transfer $\Delta C = (C_H-C_T)/2$. Indeed, the T layer can only remain a Mott insulator if the flat band is nearly half-filled, but since the work function of the H layer is larger than that of the T layer, some amount of charge transfer from T to H is expected~\cite{Friend77,Beal78,Doran78}. The Kondo effect similarly can only survive a certain charge transfer, and in addition it is only realized for a critical hybridization $V>V_c$. Superconductivity is also expected to be very different depending on whether the T layer contributes magnetic moments or not. A recent work has advocated the picture of a doped Mott insulator with negligible $V$~\cite{Crippa23}. Importantly, the effect of interlayer charge transfer in bulk 4H$_b$ compounds has not been addressed to date, nor the potential variability in the different members of the family, TaS$_2$, TaSe$_2$ or NbSe$_2$. 

In this work, we conduct a systematic study of charge transfer across all these compounds, exploring its correlation with work function mismatch, Van der Waals corrections, Hubbard's $U$ parameter and interlayer spacing. Our calculations reveal that the influence of $U$ and Van der Waals effects is minimal while we observe that charge transfer is inversely related to the distance between layers and directly linked to the mismatch in work functions. In addition, we find the general trend that Se compounds display smaller charge transfer than S compounds, and 4H$_b$ bulk polytypes display more charge transfer than isolated bilayers.

\section{Charge transfer in T/H structures}\label{ct_st}

Early works~\cite{Friend77,Beal78,Doran78} already anticipated that charge transfer from the T to the H layers must be present in bulk 4H$_b$ TMDs. At high temperatures where the CDW in the T layers is incommensurate, the change in CDW wavevector compared to bulk 1T polytypes was used to estimate a transfer of 0.12 $e^-$ per formula unit~\cite{Friend77} (1.56 $e^-$ per SoD) in 4H$_b$-TaS$_2$, and a similar estimate leads to 1.20 $e^-$ per SoD for TaSe$_2$. Similar theoretical estimates~\cite{Doran78} for 4H$_b$-TaS$_2$ similarly ranged between 1.04 to 1.43 $e^-$ per SoD. Charge transfer of this magnitude was also reported to be consistent with changes in the optical conductivity in both 4H$_b$ and 6R polytypes~\cite{Beal78}. More recent ARPES experiments estimated 0.92 $e^-$ per SoD in 4H$_b$-TaS$_2$~\cite{Almoalem24}. All these early estimates are thus consistent with a nearly empty flat band. 

Recent ab-initio calculations in the high temperature state without CDW~\cite{Yan15,Gao20} also suggest charge transfer from T to H, but given the strong band reconstruction due to the CDW, it is important to perform these calculations in the CDW state. Such calculations~\cite{Nayak21,Nayak23} for a T/H bilayer of TaS$_2$ still report a fully empty flat band, while a bilayer T on monolayer H reported 0.31 e$^-$ per SoD cell~\cite{Wang18}. For TaSe$_2$, a value of 0.32 e$^-$ per SoD cell was reported \cite{Wan22}, while for NbSe2 0.17 e$^-$ was calculated~\cite{Pico24}. A recent study has emphasized the importance of the stacking distance on charge transfer~\cite{Crippa23}, revealing that $\Delta C$ ranges from 0.4 to 1 in TaS$_2$ as the interlayer distance goes from 7 to 5.8 \AA. 

Given such variabilty, and the fact that isolated bilayers on substrates may not stack with the same interlayer distance as bulk 4H$_b$ compounds, it is important to study the distance dependence in detail for TaSe$_2$ and NbSe$_2$. Differences may be expected because the flat band in the Se compounds is significantly closer to the CDW valence bands compared to the S compounds. For NbSe$_2$ one experiment has claimed \cite{Ganguli24} a charge transfer with an opposite sign to that of the Ta compounds. It is also important to take into account the details of the different ab-initio calculations done for the T compounds~\cite{Darancet14,Zhang14,Yu17,Yi18,Chen20,Jiang21,Park20} , for example because the exact position of the flat band within the CDW gap is known to depend on the functional used \cite{Kamil18}, which can influence charge transfer. Including the Hubbard interaction and explicitly accounting for magnetic states with spin-split bands~\cite{Pasquier18,Kamil18,Tresca19,Zhang20,Pasquier21} can similarly affect the charge transfer. 
%\cite{Shao22} computed charge transfer with ice overlayers, saying the work function can be changed. 

Finally, note in the context of correlated systems the term charge transfer is often used to emphasize the distinction between Mott and charge transfer insulators~\cite{ZSA85}. In that case, the term refers to charge transfer between correlated $d$-derived bands and the dispersive $p$-derived bands of the same compound. In TMDs this phenomenon may also be relevant at least for some compounds like 1T-NbSe$_2$ or 1T-TaSe$_x$Te$_{1-x}$ where the flat band may overlap with the $p$-derived states~\cite{Kamil18,Liu21,Phillips23}. In our work, unless specified otherwise, charge transfer will rather refer to interlayer charge transfer between the T and the H layers.

\subsection{\textit{Ab-initio} methods}\label{methods_ab}

The aim of this work is to provide a systematic study of the interlayer charge transfer between T and H MX$_2$ layers. 
The workflow used to carry out said study is as follows:
\begin{enumerate}
    \item Single layer in-plane structure relaxation for H and T layers.
    \item Single layer workfunction calculation for H and T layers.
    \item T-H bilayer relaxation, first relaxing in the $\hat{z}$ direction followed by a subsequent in-plane relaxation.
    \item Charge transfer calculation for bilayers. In this step we explore different parameters as Hubbard $U$, Van der Waals corrections and inter-layer distance dependence.
    \item 4H$_b$ charge transfer calculation using step 2 structures and experimental distances.
\end{enumerate}

The purpose of this workflow is to identify trends between different factors affecting \textit{ab-initio} calculations and experimental measurements, aiming to understand the overall behaviour of charge transfer under different conditions. %The relaxation process seeks to establish a consistent basis for comparing systems, rather than estimating an accurate equilibrium distance between layers, as the interlayer distance is systematically analyzed in subsequent steps. 
All calculations were performed using Vienna Ab initio Simulation Package (VASP)~\cite{VASP1,VASP2} v.6.2.1. with projector-augmented wave pseudopotentials within the Perdew Burke Ernzerhof parametrization~\cite{Perdew96}. For step $1$ and $3$, the relaxation was conducted by using the conjugate-gradient algorithm as implemented in VASP, and setting the ionic positions as the sole degree of freedom. For steps $1-4$, calculations were found to be well converged with a $480\;eV$ kinetic cutoff and a gamma-centered $15\times15\times1$ $k-$mesh. Meanwhile, for step $5$, the self-consistent calculations were found to be well converged with a $480\;eV$ kinetic cutoff and a gamma-centered $13\times13\times3$ $k-$mesh. In step $4$, when Van der Waals corrections were considered, DFT-D3 method~\cite{Grimme2010} with zero damping was used. Also in step $4$, in order to study the effect of Coulomb repulsion, the DFT+U rotationally invariant approach~\cite{Duradev} was followed by setting different effective on-site $U$ Coulomb interactions in M's $d-$orbitals with $J=0$. These DFT+U calculations are the only collinear spin-polarized ones. The initial magnetization was set to $1.4 \mu_B$ for the central Ta/Nb atom (A atom, see Sec. \ref{distortion}), and zero for the remaining atoms. This choice is based on prior studies suggesting that the predominant magnetic moment is concentrated at the center of the SoD~\cite{Yu17}, and that the total magnetic moment is typically around $1\mu_B$~\cite{Zhang20,Pasquier18}. The initial value of magnetization is taken slightly larger than the expected result as this is expected to improve convergence \cite{vasp_magmom}. Atom-projected band structures were obtained using PyProcar~\cite{pyprocar} package for Python. 

\subsection{Charge density wave distortions}\label{distortion}

The $\sqrt{13} \times \sqrt{13}$ SoD CDW structure is shown in Fig. \ref{SoDFig}. There are three types of symmetry equivalent M sites labeled as A, B and C hereafter, with multiplicities 1, 6, and 6 respectively. The structure is parametrized by three independent displacements $\vec u_i$ with $i=A,B,C$, shown in Fig. ~\ref{SoDFig} ($|\vec u_A| = 0$ by symmetry). In table~\ref{tab:all} we present the displacements for each equivalent metal site in each 1T-MX$_2$. In appendix~\ref{app:disp} we present a graphic depiction of these displacements for the four compounds (Fig.~\ref{displacements}) along a brief note on how we obtained the CDW positions. 

%As depicted by Fig.,  each of them has their corresponding CDW displacement $\vec u_i$ with $i=A,B,C$. In table~\ref{tab:displacements} we present the averaged displacements for each equivalent metal site in each 1T-MX$_2$. This table is filled by calculating the in-plane Euclidean distance between each site in the CDW and no-CDW 1T-MX$_2$ configuration. In appendix~\ref{app:disp} we present a graphic depiction of these displacements for the four compounds (Fig.~\ref{displacements}) along a brief note on how we obtained the CDW positions. 

%\begin{table}[]
%\resizebox{0.35\textwidth}{!}{
\caption{Lattice structure of the SoD CDW distortion of the 1T polytypes. A, B, C inequivalent metal sites are marked in yellow, green and blue respectively. Black arrows show their displacements while a dotted line marks the CDW unit cell.} 
    \label{SoDFig}
\end{figure}

\subsection{Work function analysis}

We define the work function as the absolute Fermi level with respect to vacuum. Different work functions between two structures indicate misaligned Fermi levels and can be used for a qualitative estimate of charge transfer. %In this study, we aim to examine the correlation between charge transfer and the difference in work functions. 
To compute the work functions, we perform a self-consistent calculation of the monolayer in a unit cell with  $\approx 20\; {\rm\AA}$  of vacuum in the normal direction. From this calculation we extract the local potential $V(\vec{r})$ and define the work function as
\begin{align}
    W = V_{\text{vac}} - E_F,
\end{align}
where  $V_{\text{vac}}$ is the value of $V(\vec{r})$ in vaccuum. 
We present the work functions for both T and H polytypes of all TMDs considered in table~\ref{tab:all}. H work function is greater than T work function for all TMDs which anticipates that the charge transfer will occur from T to H layer. In Table~\ref{tab:all} 
we can already see some trends: M=Ta compounds have overall smaller work function than M=Nb and so happens with X=Se when compared with X=S ones. In section~\ref{CT} we will see how this affect the charge transfer. 

% \begin{table}[]
% \resizebox{0.4\textwidth}{!}{
% \begin{tabular}{l|cccc|l}
% \cline{2-5}
%                         & NbSe$_2$ & NbS$_2$ & TaSe$_2$ & TaS$_2$ &  \\ \cline{1-5}
% \multicolumn{1}{|c|}{$W_{\text{1T}}$ (eV)} & 5.25  & 5.47 & 5.02  & 5.19 &  \\ \cline{1-5}
% \multicolumn{1}{|c|}{$W_{\text{1H}}$ (eV)} & 5.57  & 6.13 & 5.45  & 5.57 &  \\ \cline{1-5}
% \end{tabular}}
% \caption{Work functions for the 1T and 1H configurations in all four TMDs considered}
% \label{tab:workf}
% \end{table}

\begin{table*}
\begin{tabular*}{1.5\columnwidth}{@{\extracolsep{\stretch{1}}}*{5}{c}@{}}
  \toprule
  & NbSe$_2$ & NbS$_2$ & TaSe$_2$ & TaS$_2$ \\
    \hline
  $|\vec{u}_A|$ (\AA) & 0.00 & 0.00 & 0.00 & 0.00   \\
  $|\vec{u}_B|$ (\AA) & 0.26 & 0.21 & 0.26 & 0.20   \\
  $|\vec{u}_C|$ (\AA) & 0.32 & 0.28 & 0.31 & 0.25  \\
  $d$ (\AA) & 7.37 & 6.89 & 7.51 & 7.10  \\
  $W_{T}$ (eV)  & 5.25 & 5.47 & 5.02 & 5.19   \\
  $W_{H}$ (eV)  & 5.57 & 6.13 & 5.45 & 5.57 \\
  $\Delta$WF (eV)  & 0.32 & 0.66 & 0.43 & 0.77   \\
  CT (e)  & 0.12 & 0.23 & 0.14 & 0.26   \\
  \hline                             
\end{tabular*}
\caption{Displacements for each inequivalent metallic position $\vec{u}_i$ with $i=A,B,C$, interlayer distance $d$, work functions $W_T$ and $W_H$ for T and H structures, workfunction differences $\Delta$WF and charge transfer CT for all four compounds considered.}
\label{tab:all}
\end{table*}

The work function results in Table~\ref{tab:all} are consistent with Ref.~\cite{Wang18}, which reported work function values of $W = 5.35$ eV for T-TaS$_2$ in the CDW state, and $W = 6.07$ eV for H-TaS$_2$ in the $3\times3$ CDW state. Here we calculated $W = 5.19$ eV for T-TaS$_2$ in the CDW state, and $W = 5.57$ eV for 1H-TaS$_2$ without CDW. We opted not to include the $3\times3$ CDW state in H-TaS$_2$ layers in our calculations due to the excessive computational cost of a unit cell commensurate with both $\sqrt{13} \times \sqrt{13}$ and $3\times3$ CDW states. However, our analysis of the charge transfer as function of work function differences suggests this is a good approximation.

\subsection{Charge transfer}\label{CT}

\begin{figure}[t]
    \centering
\includegraphics[width=0.45\textwidth]{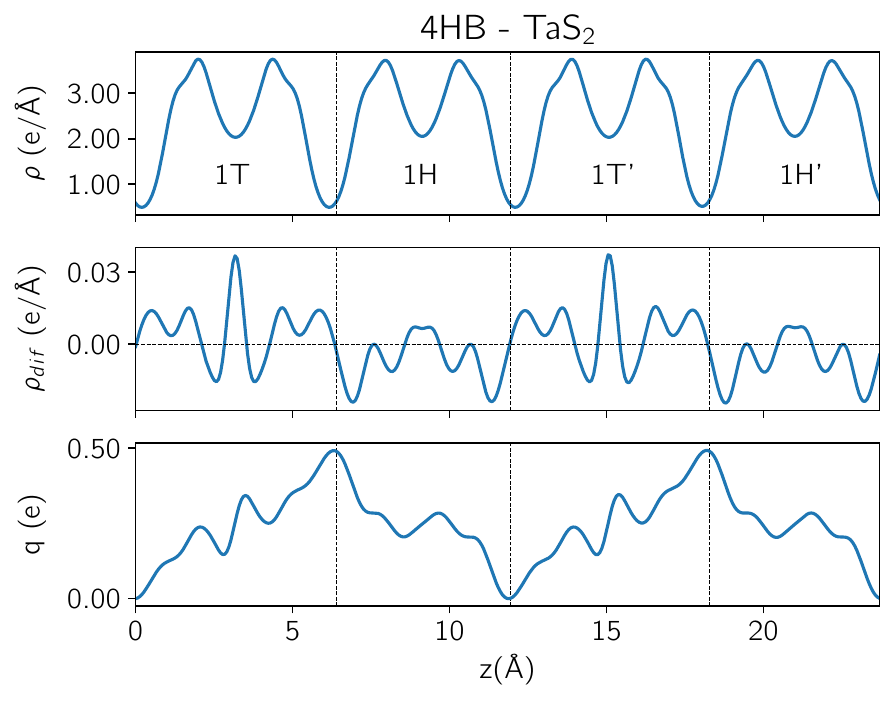}
     \caption{Electronic density (top), electronic density difference (mid) and total charge change (bottom) for 4H$_b$ TaS$_2$. Zeros for electronic density difference and thus candidates for $z_1/z_2$ are marked with vertical dashed lines.}
    \label{4H$_b$}
\end{figure}

\begin{figure}[t]
    \centering
\includegraphics[width=0.45\textwidth]{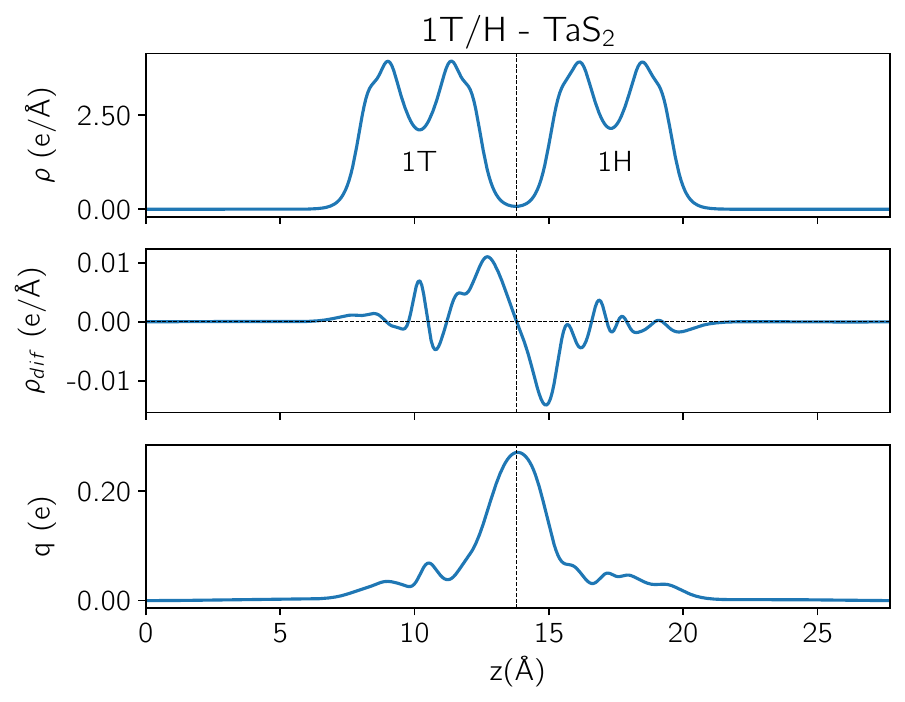}
     \caption{Electronic density (top), electronic density difference (mid) and total charge change (bottom) for 1T/H TaS$_2$ bilayer. The zero for electronic density difference is marked with a vertical dashed line.}
    \label{1t_ct}
\end{figure}

Charge transfer is the central result of this work. To calculate the charge transfer between the different constituents of a heterostructure we followed the method presented in Refs.\cite{Wang18,ct2}. This method is based on computing the charge density of the full structure with respect to that of a hypothetical reference structure built from the calculated charge densities of the isolated constituents, positioned in the places they would occupy in the full structure. More explicitly, if we consider $\hat{z}$ as the stacking direction,  we can obtain the plane-averaged charge density from the self-consistent calculations $\rho_{\text{all}}(z)\;$ and $\rho_{\text{i}}(z)$ with $i=1,2, ..., N$, where $N$ is the number of component, and obtain the overall charge density difference as 
    \begin{align}
        \rho_{\text{dif}}(z) = \rho_{\text{all}}(z) - \sum_{i=1}^N \rho_i(z),
    \end{align}
Integrating $\rho_{\text{dif}}(z)$ we can obtain the total charge difference in a section $(z_1,z_2)$ as 
    \begin{align}
        q_{\text{dif}}(z_1,z_2) = \int_{z_1}^{z_2} \rho_{\text{dif}}(z) \text{d}z.
    \end{align}
Determining appropriate values for $z_1$ and $z_2$ can be challenging when $N>2$ or for periodic systems, such as the 4H$_b$ structure. A practical approach is to set $z_1$ as the point where $\rho_{\text{dif}}(z_1)=0$ nearest to the interface with the preceding component, and $z_2$ as the point where $\rho_{\text{dif}}(z_2)=0$ nearest to the interface with the subsequent component. For instance, in Fig. \ref{4H$_b$}, the total charge difference for the T layer in 4H$_b$ is obtained by integrating from $z_1=0\;$\AA to $z_2 = 6.40\;$\AA~ (where the vertical and horizontal dashed lines first intersect in the second graph); for H, integration ranges from $z'_1=z_2=6.40\;$\AA~ to $z'_2 = 11.95\;$\AA~, and so forth. These integration limits correspond to the maximum and minimum values of $q(0,z)$ as illustrated in the third graph. In the scenario of $N=2$ isolated components, integration simplifies: as depicted in Fig.~\ref{1t_ct}, $z_1$ may be positioned anywhere in vacuum, while $z_2=2$ marks the point where $\rho_{\text{dif}}(z_2)=0$ nearest to the interface. $q(z_1,z_2)$ represents the total charge difference for the first component and the overall absolute charge transfer between components.

\begin{figure}[t]
\vspace*{-0.7cm}
\hspace*{-0.4cm}
    \centering
\includegraphics[width=0.55\textwidth]{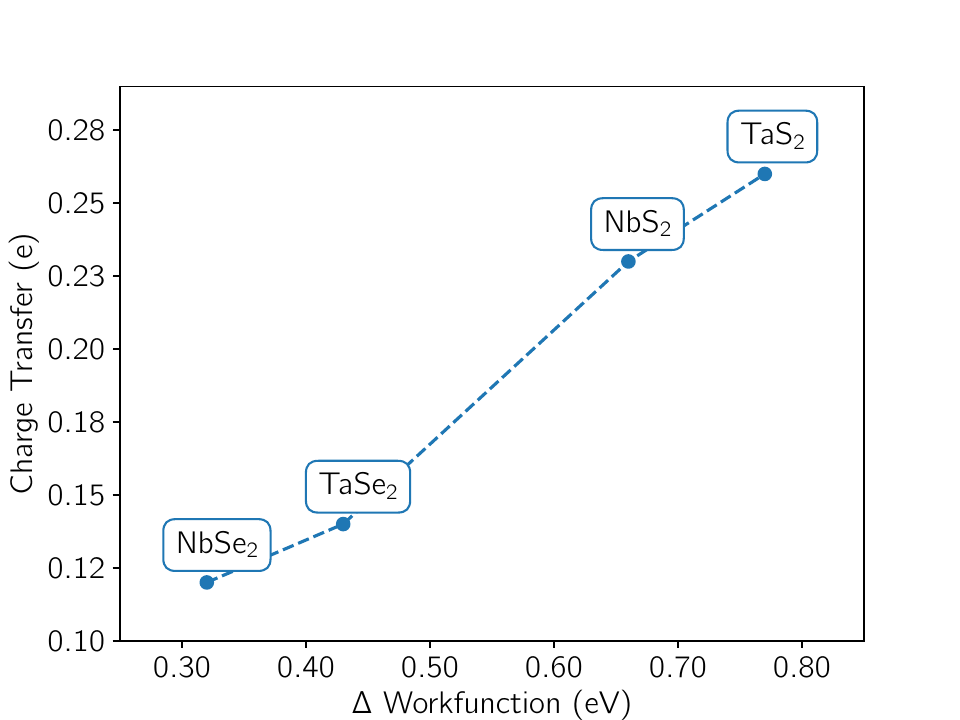}
     \caption{Charge transfer as a function of workfunction difference between layers in 1T/H bilayer.}
    \label{wf_vs_ct}
\end{figure}

In Table~\ref{tab:all} we present the charge transfer results for the relaxed interlayer distances. These distances are $6.89\text{\AA}$ for NbS$_2$, $7.37\text{\AA}$ for NbSe$_2$, $7.10\text{\AA}$ for TaS$_2$ and $7.51\text{\AA}$ for TaSe$_2$. Those values are plotted in Fig.~\ref{wf_vs_ct}, where the previously mentioned $CT\propto \Delta WF$ trend is clear. From these results, we can  detect some trends: \textit{ab-inito} calculations predict Ta compounds to have greater charge transfer than Nb compounds, and S compounds have greater charge transfer than Se compounds. 

%Integrate the electronic density difference along the normal direction starting from a point of zero charge density $\rho_{\text{dif}} = \rho_{1T/H} - (\rho_{1T} + \rho_{1H})$. This integrated value at the will peak 
%When working with two isolated monolayers, 
%For an isolated 1T/1H-TaSe$_2$ bilayer with the T layer in the CDW state, 0.20 without SOC, 0.18 with SOC. When the T layer is not in the CDW state, charge transfer is actually much higher, 0.49. So it matters a lot to have the CDW.
%For an isolated 1T/1H-TaS$_2$ bilayer with the T layer in the CDW state, 0.31 without SOC.
%4H$_b$-TaS$_2$, with relaxation of atomic positions and lattice constants (but preserving volume) yields 0.46 without SOC and 0.45 with SOC, per bilayer (unit cell contains two bilayers). 

\subsection{Van der Waals effect}\label{vdW}

In this section, we compare the calculation of the charge transfer with and without 
Van der Waals corrections for the T/H structures obtained in Sec.~\ref{CT}. To do so we incorporated Van der Waals corrections into the electronic self-consistent calculations and repeated the charge transfer calculations in Sec. \ref{CT}. Our findings indicate that these corrections have a negligible influence on charge transfer, typically on the order of $<10^{-3}$. 

While Van der Waals corrections therefore do not affect charge transfer directly, the could do so indirectly if we performed a new relaxation of the structure in the presence of such corrections, since the interlayer distance could change upon relaxation. Since our main interest in this work is to establish relative trends in charge transfer, we have rather opted to study charge transfer as a function of interlayer distance without attempting to calculate precisely its equilibrium value, as this is a more complex problem that depends on both calculational details and experimental conditions. 

%It is noteworthy to mention that, while these corrections may become significant when integrated into relaxation calculations, we opted to exclude them from our methodology because we believe that it adequately accounts for the prospective corrections in interlayer distances by performing an extensive exploration on distance dependence. 

\subsection{U dependence}

%In accordance with the procedure followed in Sec.~\ref{vdW} for Van der Waals corrections, electron-electron interactions were exclusively accounted for in electronic self-consistent calculations for the relaxed distances \noteIS{but solely} in TaS$_2$. \noteMV{no me debo estar enterando bien, lo de VdW no es general? } \noteIS{No, está mal explicado o poco claro?}}

\begin{figure}[t]
    \centering
    \hspace*{-.5cm}
\includegraphics[width=0.4\textwidth]{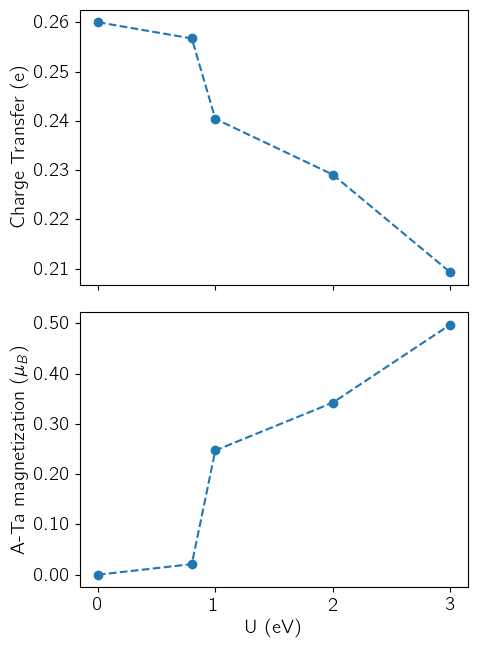}
     \caption{Charge transfer (top) and magnetization of A atoms in the T layer (bottom) dependence on U for TaS$_2$.}
    \label{UFig}
\end{figure}

Following the same logic as in the previous section, we now consider the effect of the Hubbard interaction $U$, only at the level of electronic self-consistent calculations keeping the structure fixed. We considered only TaS$_2$ as an example. The main effect of the Hubbard interaction is to magnetize the flat band near the Fermi level, producing a spin splitting and pushing one of the spin polarizations above the Fermi level. Since a larger magnetization is produced when the lower spin-split band is closer to half-filling, increasing $U$ generally leads to a reduction in charge transfer to gain energy from magnetization. The effect should thus be larger in the compound with larger charge transfer, which is TaS$_2$. Figure \ref{UFig} illustrates both charge transfer and magnetization of the center of the SoD as a function of $U$ for this compound. As expected, charge transfer exhibits an inverse relationship with $U$, albeit the overall impact of $U$ on charge transfer remains relatively minor (on the order of $10^{-2}$e). The critical $U$ for the flat band magnetization is around $U_c \sim 1$ eV, as we can see from the spin split bands in Fig. \ref{app:bands}, where we also include a brief comment on the effect of $U$ to the electronic band structures. The conclusion of this section is that while $U$ might have a sizable effect in the bands and magnetization, is not so relevant for charge transfer.

%To this end, four distinct values were considered for $U = (0.8,1.0,2.0,3.0)$~eV, initiating the self-consistent calculation with a magnetic moment of $1.4\mu_B$ at the center of the SoD (A M's atoms). 

%This decision stems from two primary considerations. Firstly, the anticipated consistency of $U$'s effect across all compounds, which has been determined to be nearly negligible. S

%econdly, given that $U$ induces a splitting of the 1T's flat band into two spinful bands, with one being pushed up into the valence bands, it is anticipated that charge transfer will exhibit an inverse relationship with $U$ \noteMV{acabas de decir que el effect es nearly negligible} \noteIS{El efecto es muy pequeño para la CT, del orden de 0.01 e/u.c, pero el efecto en las bandas es el descrito arriba}. Thus, a greater extent of charge transfer is expected to provide a more suitable basis for investigating the effects of $U$. 

\subsection{Distance dependence}
%In order to understand the role of interlayer distance in charge transfer we repeated step 3 from Sec.~\ref{methods_ab} for various distances, including both the relaxed and \textit{experimental} values. \textit{Experimental} values are obtained from the reported distance between layers in bulk 4H$_b$ as follows. The out-of-plane lattice constant for TaS$_2$ is $c=23.73 \; {\rm \AA}$ \cite{Ribak20}, and hence the Ta-Ta interlayer distance is $5.93 \; {\rm \AA}$. For TaSe$_2$ is $c=25.16 \; {\rm \AA}$ \cite{Morbt79}, so the Ta-Ta interlayer distance is $6.29 \; {\rm \AA}$. For NbSe$_2$ the 4H$_b$ phase has never been reported, but a T-H interlayer distance can be extrapolated as follows: the different NbSe$_2$ polytypes composed only of H layers all have interlayer distances of $6.27-6.31 \; {\rm \AA}$, while the analog polytypes for TaSe$_2$ have interlayer distances $6.35-6.39  \; {\rm \AA}$ \cite{Kalikhman73}. Taking the lowest value as a reference, the interlayer distance for 4H$_b$-TaSe$_2$ is only 1\% lower than the one for the H polytypes, so we can extrapolate a 1\% reduction for a hypotethical 4H$_b$-NbSe$_2$ and assume an interlayer distance of $6.21  \; {\rm \AA}$. NbS$_2$ is not studied because it is not experimentally reported.  

In order to explore the impact of interlayer distance on charge transfer, we calculated the charge transfer across a range of distances for experimentally reported compounds, encompassing both the relaxed values and those derived from experimental data. The structure of each layer, obtained after relaxation, is kept fixed as the interlayer distance is varied, as we expected the effect of further relaxation to be minimal except perhaps at very close distances. Experimental values were determined based on the reported interlayer spacings in bulk 4H$_b$ materials. Specifically, for TaS$_2$ ($c=23.73 {\rm \AA}$) \cite{Ribak20}, the Ta-Ta interlayer distance is $5.93\; {\rm \AA}$. Likewise, for TaSe$_2$ ($c=25.16\;{\rm \AA}$) \cite{Morbt79}, the corresponding interlayer distance is $6.29\; {\rm \AA}$. For NbSe$_2$, the 4H$_b$ polytype has not been reported, so we extrapolate its interlayer distance by comparison with TaSe$_2$. The interlayer distance in 2H-TaSe$_2$ is $6.35 \; {\rm \AA}$, which is 1\% less than that of 4H$_b$-TaSe$_2$~\cite{Kalikhman73}. For 2H-NbSe$_2$ the interlayer distance is $6.27 \; {\rm \AA}$, so assuming the same trend we can take 1\% less for a hypothetical 4H$_b$-NbSe$_2$ structure, i.e. $6.21$. NbS$_2$ was not included in this analysis due to lack of experimental data. It should be noted that the experimental interlayer distances in bulk compounds need not be the same as for bilayers on a substrate, so these values should be taken only as reference. In general, we also see that these values are smaller than those obtained from relaxation in Table \ref{tab:all}. 

%by comparing the polytypes composed solely of H layers of NbSe$_2$ and TaSe$_2$, which exhibit interlayer distances of $6.27-6.31 \; {\rm \AA}$ and $6.35-6.39 \; {\rm \AA}$ respectively \cite{Kalikhman73}. Considering the lower end of this range as a reference, the interlayer distance for 4H$_b$-TaSe$_2$ is only 1\% less than that of the corresponding H polytype. Using the same factor we obtain an extrapolated interlayer distance of $6.21 \; {\rm \AA}$ for a hypothetical 4H$_b$-NbSe$_2$. NbS$_2$ was not included in this analysis due to lack of experimental data.

%The dependence on interlayer distance for the three compunds NbSe$_2$, TaS$_2$ and TaSe$_2$ is shown in Fig. \ref{DistanceFig}. It is noticeable that all three compounds present a similar charge transfer dependence on distance, presenting two different patterns: for greater distances, there is a slow linear growth of charge transfer with distance, while for smaller distances, after a tipping point the linear growth gets steeper. 

\begin{figure}[]
    \centering
\includegraphics[width=0.48\textwidth]{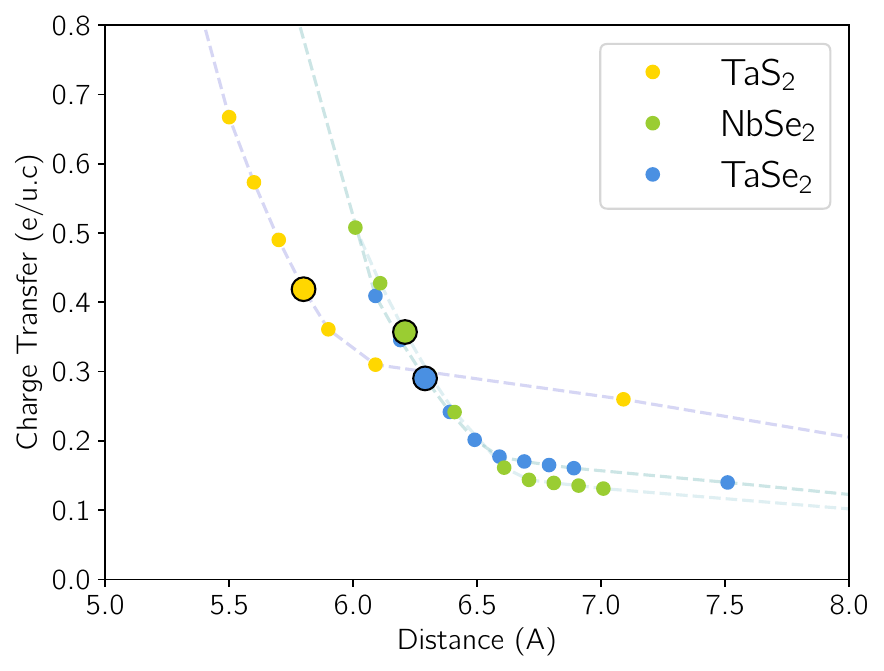}
     \caption{Charge transfer dependence on interlayer distance for T/H bilayers of NbSe$_2$, TaS$_2$ and TaSe$_2$ in the non-magnetic state at $U=0$. Experimental equilibirum distances for 4H$_b$ compunds marked by a bigger dot with a black contour (for NbSe$_2$ an extrapolated value is used, see text). }
    \label{DistanceFig}
\end{figure}

Figure \ref{DistanceFig} illustrates the interlayer distance dependence of charge transfer for NbSe$_2$, TaS$_2$, and TaSe$_2$. All three compounds exhibit a comparable trend in charge transfer as a function of distance, characterized by two distinct patterns: at larger interlayer distances, a gradual linear increase in charge transfer is observed, while at smaller distances, a tipping point is reached, beyond which the rate of linear growth becomes more pronounced.  From distances of 5.5 \AA~and below, our charge transfer calculation method becomes less reliable due to the ambiguity in defining the boundaries between subsystems in the presence of strong hybridization. We also observe that TaS$_2$ has the largest charge transfer also at the experimental interlayer distance. 

%ineffective due to hybridization, even suggesting a charge transfer exceeding $1$ e/u.c for NbSe$_2$ and TaS$_2$. 

%In order to place our results within previous works we can now compare them by extrapolating to different distances. For example, the results presented in Ref. \cite{Wan22} show a charge transfer of 0.32 $e$ with an interlayer distance of 6.6 \AA~ for TaSe$_2$, which is greater than our result. Meanwhile, the work on NbSe$_2$ in Ref.\cite{Pico24} where the interlayer distance is 6.08 \AA~ shows a charge transfer of 0.17 $e$, smaller than our predictions of 0.43 $e$ presented in Fig.~\ref{DistanceFig}. Ref~\cite{Crippa23} conducts a similar calculation for TaS$_2$ but predicts greater charge transfers overall, with a consistent growth until total charge transfer rather than a two step dependence as our results. This discrepancy may stem from variations in the charge transfer calculation methods employed but could potentially obscure the proposed image of the doped Mott insulator.

In order to contextualize our findings within existing literature, we can now compare them by extrapolating to various distances. For instance, the results reported in Ref. \cite{Wan22} predict a charge transfer of 0.32 $e$ for TaSe$_2$ at an interlayer distance of 6.60 \AA, which exceeds our result of around 0.20 $e$. Conversely, the investigation conducted on NbSe$_2$ in Ref. \cite{Pico24}, where the interlayer distance is 6.08 \AA, indicates a charge transfer of 0.17 $e$, lower than our predicted value of 0.43 $e$. %These results can be compared by choosing a distance and intersecting a vertical line with the charge transfer vs. distance curves at Fig.~\ref{DistanceFig}. 
Ref. \cite{Crippa23} undertakes a similar analysis for TaS$_2$ but predicts higher overall charge transfers and a smoother evolution towards CT=1 at small distances, rather than the two-step dependence observed in our results. This disparity may arise from differences in the methods used for calculating charge transfer as well as in the relaxation procedure, since Ref.~\cite{Crippa23} performs a relaxation for each distance, which becomes progressively more important as interlayer hybridization increases. Our work is however generally consistent with the fact that interlayer distance is the key parameter that most strongly affects charge transfer, as put forward in Ref.~\cite{Crippa23}.

\section{Results: Bulk 4H$_b$ polytypes}

To compare with the results obtained from bilayers, we calculated the charge transfer in 4H$_b$-TaS$_2$ and 4H$_b$-TaSe$_2$. We constructed the lattice structure in the CDW state by stacking the relaxed structures obtained from single layers in Sec. \ref{methods_ab} using experimentally reported interlayer distances.

\begin{figure*}[t]
    \centering
\includegraphics[width=0.48\textwidth]{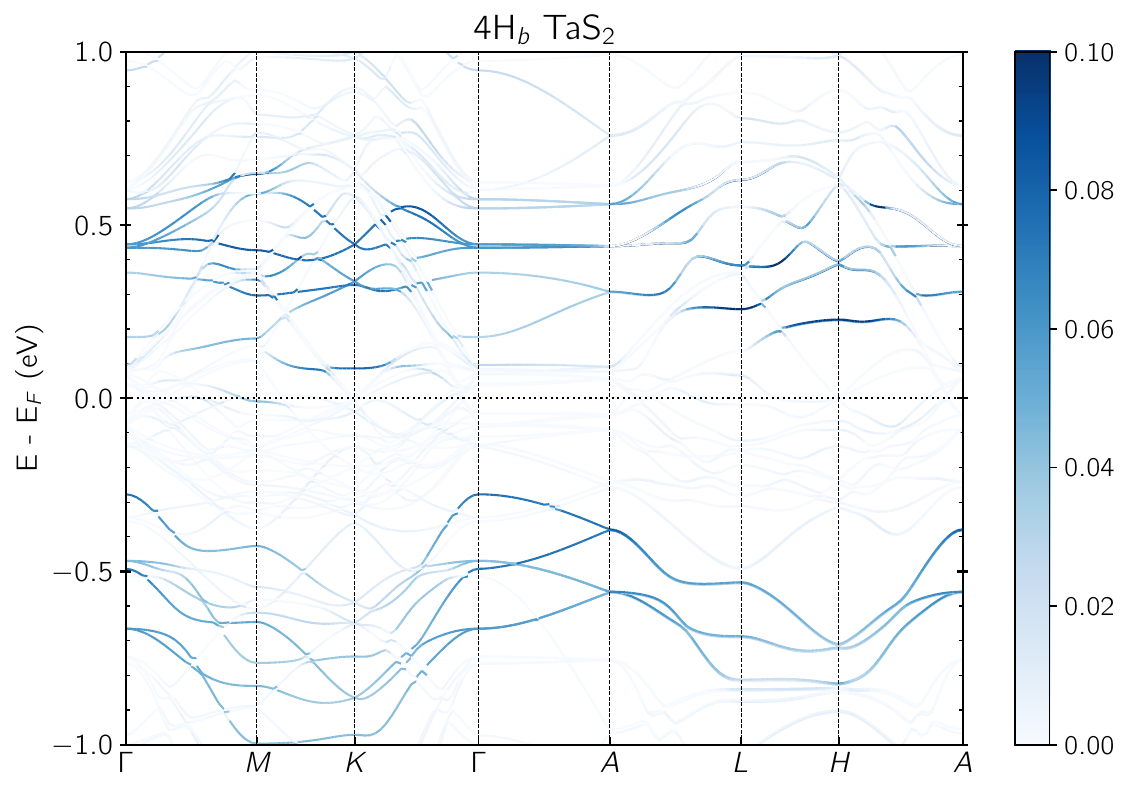}
\includegraphics[width=0.48\textwidth]{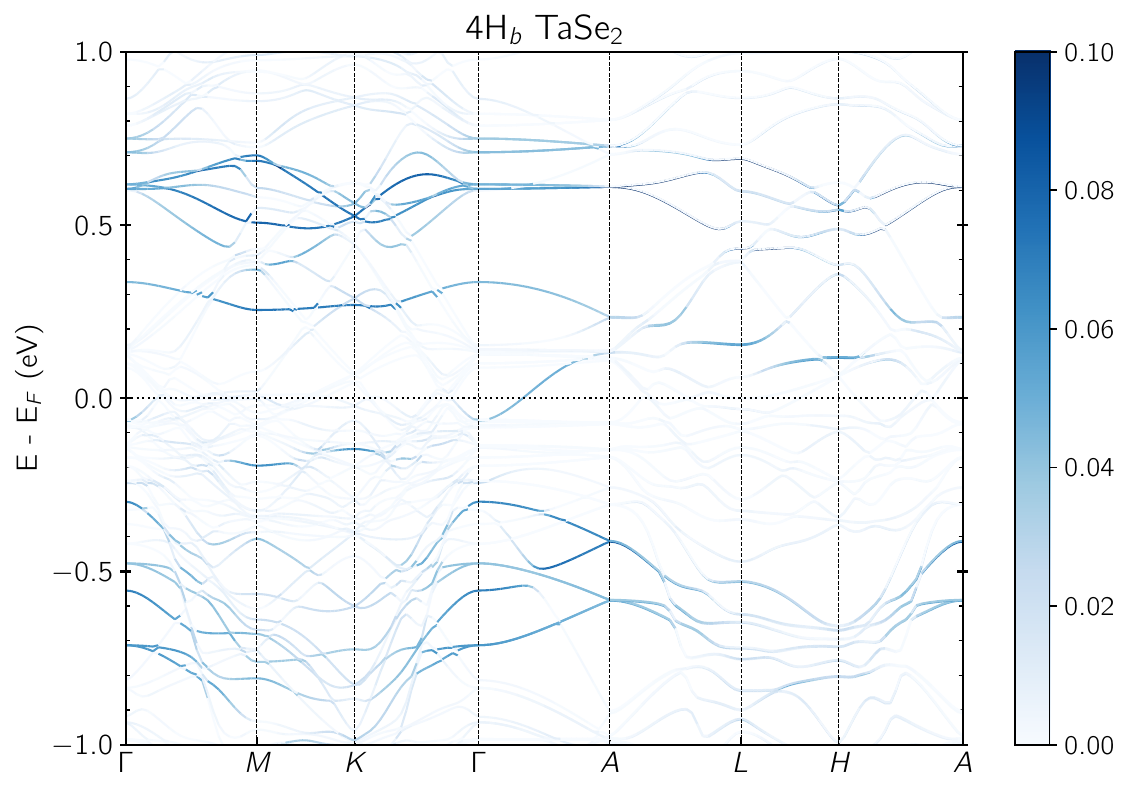}
     \caption{Band structures for 4H$_b$-TaS$_2$ (left) and 4H$_b$-TaSe$_2$ (right) in the structure when the 1T layer is in the CDW state, color coded by the weight of the wave function in the central SoD atom A. The SoD centers for both T layers are on top of each other. }
    \label{4hb_bands}
\end{figure*}

\begin{figure*}[t]
    \centering
\includegraphics[width=0.49\textwidth]{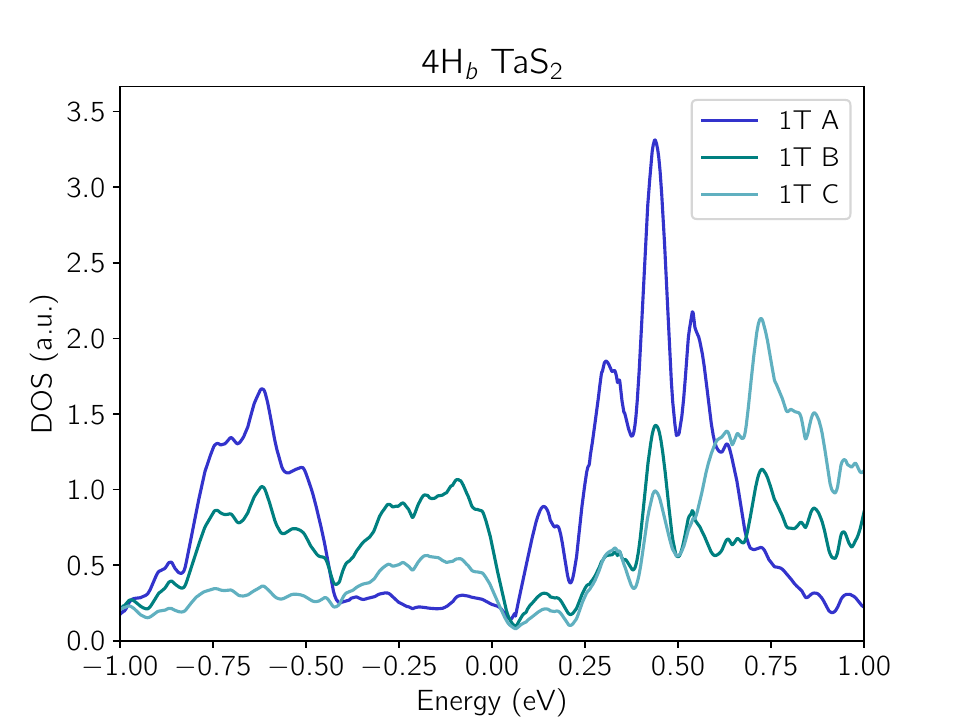}
\includegraphics[width=0.49\textwidth]{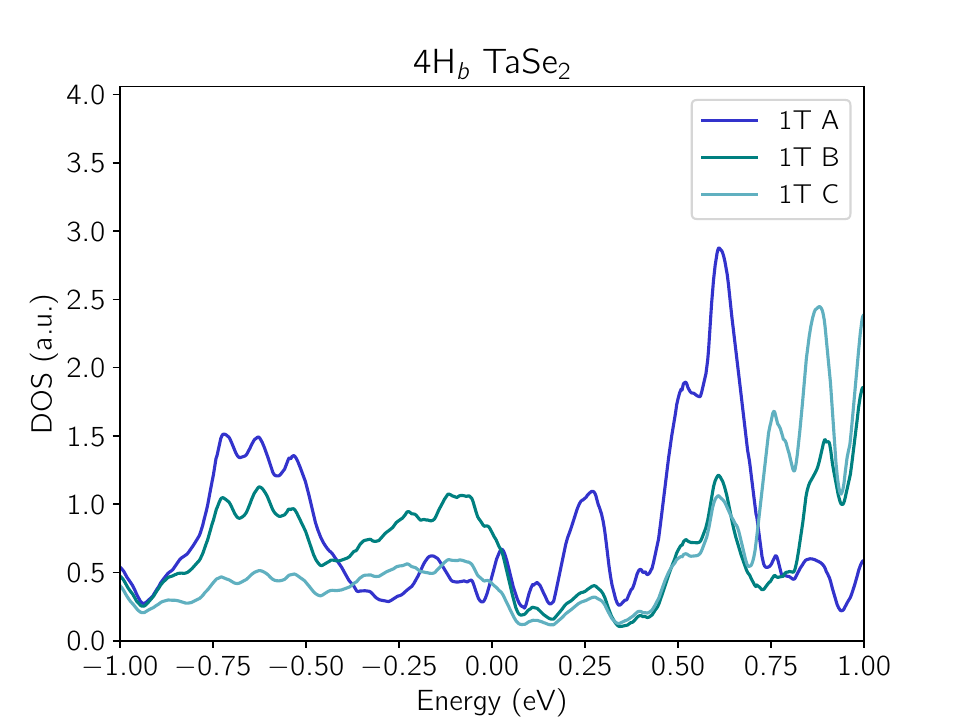}
     \caption{A,B and C atom-projected DOS for 1T-TaS$_2$ (left) and 1T-TaSe$_2$ (right) in the 4H$_b$  structure when the 1T layer is in the CDW state. }
    \label{4hb_dos}
\end{figure*}

 The charge transfer results for 4H$_b$ structures of TaS$_2$ and TaSe$_2$ compounds are $0.49$ e/u.c. and $0.39$ e/u.c., respectively. These outcomes exhibit consistency with those reported for bilayers and their distance dependence in the preceding sections. To understand these results we also computed the band structure projected to the A atom of the SoD in Fig. \ref{4hb_bands}, as well as the partial DOS calculations for A,B and C atoms in Fig. \ref{4hb_dos}, for the 4H$_b$ structures of both compounds. Interestingly, TaS$_2$ displays a flat band which is nearly everywhere above the Fermi level, consistent with a peak in the DOS for the A atom in the range 0-0.4 eV. Taken at face value this figure would suggest a charge transfer of nearly 1 e/u.c. Similarly, in TaSe$_2$, the flat bands begin at $E\approx 0.3$ eV and $E\approx -0.2$ eV, with the latter seeming to hybridize and reach $E\approx 0.1$ eV after $\Gamma$, where it exhibits a higher DOS, as illustrated in Fig. \ref{4hb_dos}. Charge transfer is harder to quantify in this case but it also appears larger than the computed value of 0.39. These disparities lead us to consider two potential explanations: firstly, that the behaviour of bands away from the high-symmetry $k$-path has a significant influence on charge transfer, as there might be bands with significant weight in the A atom below the Fermi level we simply do not see in the high symmetry path. Alternatively, interlayer hybridization might already strong enough that charge transfer cannot be inferred directly from the population of the projected bands, and the computed numerical values should be taken as the reliable ones.

%\section{Magnetism and charge transfer}

%We consider the effect of $U$. In general, charge transfer and magnetism influence each other and cannot be studied separately. 

%\section{Optics}

%The optical conductivity of 4H$_b$ and 6R TaS$_2$ show a peak around 3.2 eV~\cite{Beal78}, which is at lower energies than the peak found in 2H-TaS$_2$ (or the predicted peak for 4H$_b$ just averaging the known T and H conductivities). The old papers say this comes from charge transfer. Can we compute the optical conductivity and make a statement about charge transfer from the position of the peak?

\section{Conclusion}

%In this work we present a detailed study of charge transfer in 1T/H bilayers of MX$_2$ and 4H$_b$ TaX$_2$ structures (M=Nb,Ta, X=S, Se). This study unravels the relative magnitude of charge transfer in these compounds along its dependence on Hubbard's $U$ and interlayer distance, most important results are present in Table~\ref{tab:all} and Fig.~\ref{DistanceFig}. As a general rule of thumb, X=Se compounds present greater charge transfer than those of X=S, while M=Ta compounds present also a greater charge transfer when compared with those of M=Nb. Charge transfer is also predicted to be greater in the case of bulk 4H$_b$ when compared with bilayers but DOS and projected bandstructures do not match with this image, rendering our method useless for 4H$_b$ bulk charge transfer calculation. On the one hand, our results suggest that $U$ and Van der Waals corrections have an overall negligible effect on charge transfer. We observe that greater $U$ leads to a greater magnetization which is completely located in M's A atoms and hinders charge transfer by pushing one of the non degenerate flat bands up in energy far from Fermi level. On the other hand, our calculations predicts distance between layers to be an important factor in charge transfer. Smaller distances mean greater charge transfers and the overall trend for same X element compounds is similar: a slow somewhat linear increase of charge transfer until a tipping point after which the linear increase becomes greater until charge transfer reaches almost 1 e/u.c.

In this work, we conduct detailed \textit{ab-initio} study of charge transfer in 1T/H bilayers of MX$_2$ and 4H$_b$ TaX$_2$ structures (M=Nb,Ta, X=S, Se), revealing its dependence on Hubbard $U$, Van der Waals corrections and interlayer distance. The most significant findings are summarized in Table~\ref{tab:all} and illustrated in Fig.~\ref{DistanceFig}. Our results indicate that both $U$ and Van der Waals corrections have minimal impact on charge transfer, despite a finite magnetization. In line with previous work for TaS$_2$ \cite{Crippa23}, we find the interlayer distance to be the parameter on which charge transfer is most strongly dependent. 

The main result of our work is to reveal the different trends across T/H compounds. Generally, X=Se compounds exhibit higher charge transfer compared to those with X=S, while M=Ta compounds also demonstrate greater charge transfer compared to M=Nb counterparts. Additionally, our method predicts charge transfer to be more pronounced in bulk 4H$_b$ structures when compared to bilayers. These findings may have important experimental implications. First, there are clear differences in the behaviour of the Kondo peak in T/H bilayers in TaSe$_2$~\cite{Ruan21,Wan22}, TaS$_2$~\cite{Vavno21,Ayani22} and NbSe$_2$~\cite{Liu21,Ganguli24}, as well as in the orbital character of the higher energy states \cite{Ruan21}. Rather than inconsistencies between experiments, these features may reveal that charge transfer and the proximity to the potential Mott insulator state may be different across compounds, and more work is needed to determine charge transfer for each of them experimentally. Similarly, 4H$_b$ TaS$_2$ appears to have larger CT with more clear evidence of nearly unoccupied flat band~\cite{Nayak21,Shen22,Nayak23} which plays a minor role in superconductivity, but our calculations suggest the flat band might be more populated in 4H$_b$ TaSe$_2$ and have a stronger effect on the superconducting state \cite{Liu14Coexistence,Xie23,Liu23Existence}. We hope that our results will stimulate further work to experimentally map out charge transfer accross the family of T/H structures, which will lead to a deeper understanding of the unconventional magnetic and superconducting properties in this family of materials. 

%These findings contribute to the theoretical understanding of TMDs and represent a progression towards establishing a comprehensive framework for their computational and experimental investigation. By highlighting trends and computational methodologies, these results pave the way for a deeper comprehension of the underlying physics of TMDs.

\section{Acknowledgements} We acknowledge useful discussions with M. Ugeda and M. Gastiasoro. F. J. is supported by Grant PID2021-128760NB0-I00 from the Spanish MCIN/AEI/10.13039/501100011033/FEDER, EU. M.G.V. and I.S. thank support from the Deutsche Forschungsgemeinschaft (DFG, German Research Foundation) GA3314/1-1 -FOR 5249 (QUAST) and to the Spanish Ministerio de Ciencia e Innovacion grant PID2022-142008NB-I0. M.G.V. acknowledges partial support from the European Research Council (ERC) under grant agreement no. 101020833.

\appendix
\section{Displacements}\label{app:disp}
\begin{figure*}[t]
    \centering
\includegraphics[width=0.45\textwidth]{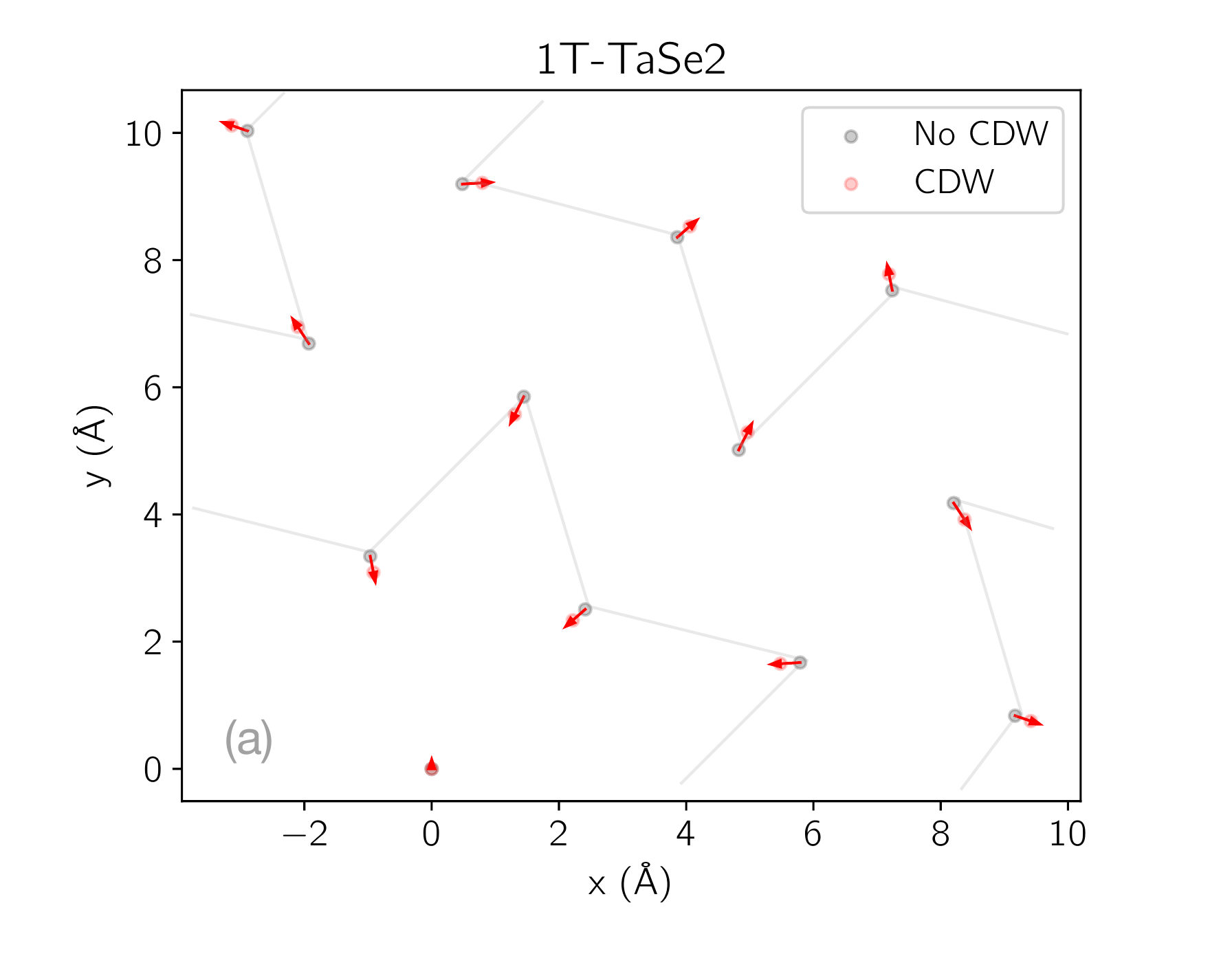}
\includegraphics[width=0.45\textwidth]{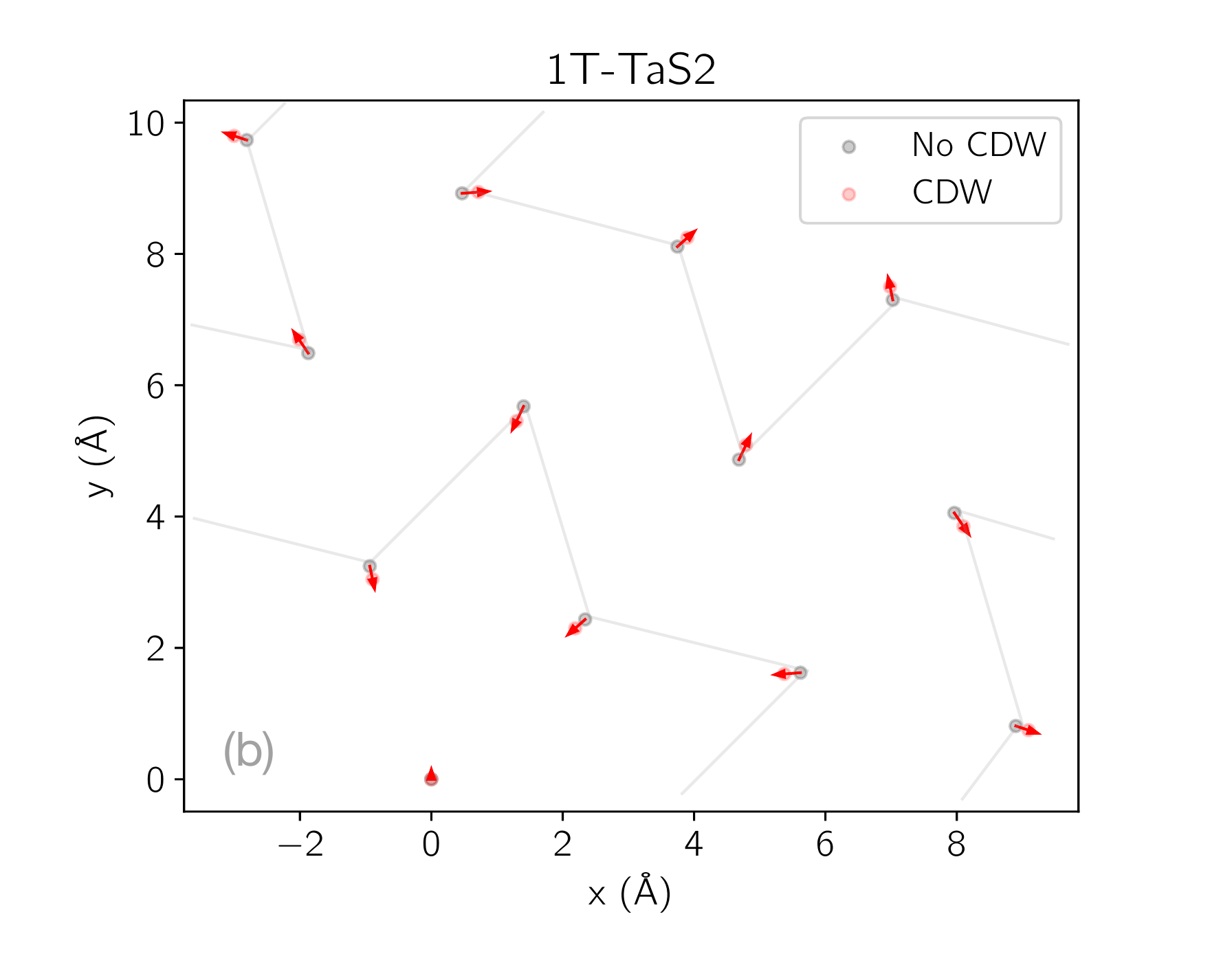}
\includegraphics[width=0.45\textwidth]{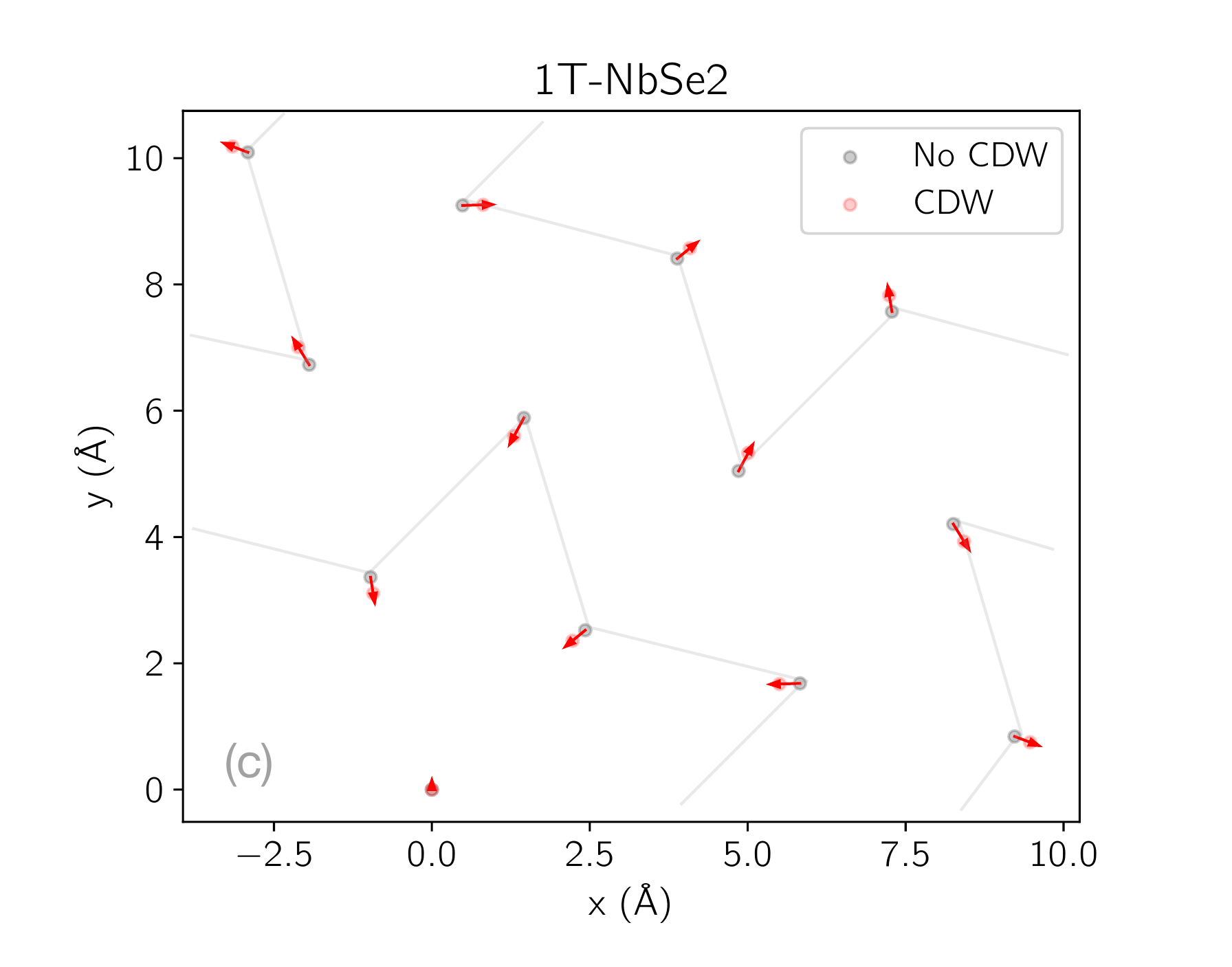}
\includegraphics[width=0.45\textwidth]{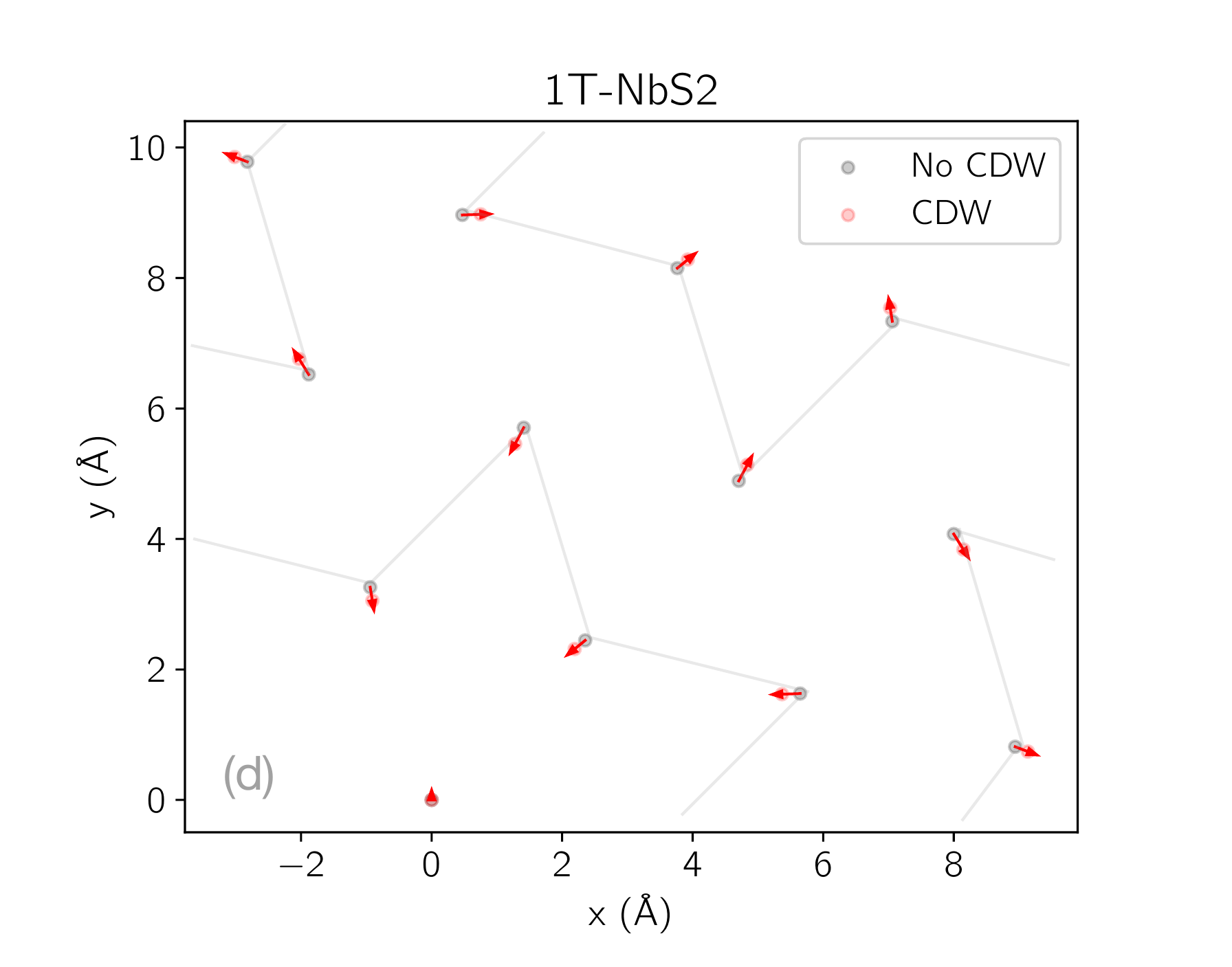}
     \caption{ CDW and no CDW atomic positions and displacements of (a) 1T-TaSe$_2$ (b) 1T-TaS$_2$ (c) NbSe$_2$ and (d) 1T-NbS$_2$. SoD CDW its superimposed in order to show the different symmetry positions A,B and C.}
    \label{displacements}
\end{figure*}

We present the positions for CDW and no CDW configurations along the displacements for all four TMDs in Fig.~\ref{displacements}. All relaxed CDW systems were obtained from a calculation whose initial state had small displacements following the trend of our previous work~\cite{Wan22}. All VASP's POSCARS used and obtained are present in supplementary files. 

\section{U bands}\label{app:bands}
\begin{figure*}[t]
    \centering
\includegraphics[width=0.48\textwidth]{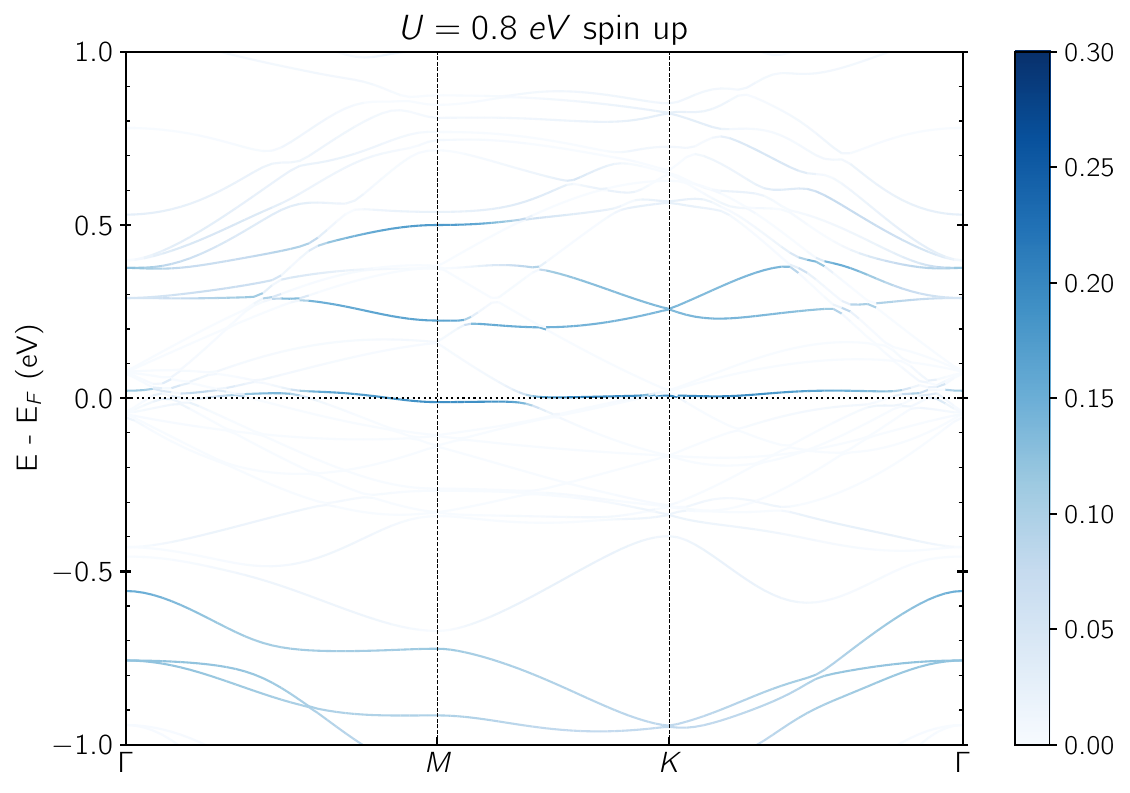}
\includegraphics[width=0.48\textwidth]{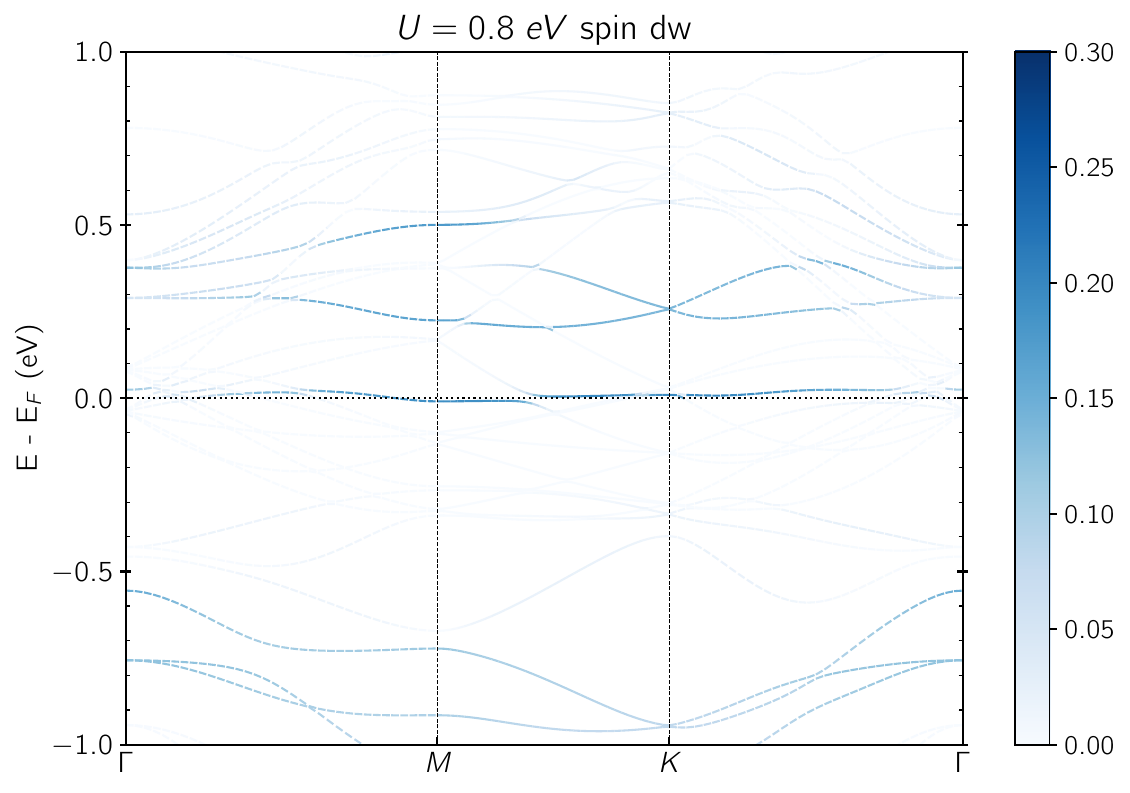}
\includegraphics[width=0.48\textwidth]{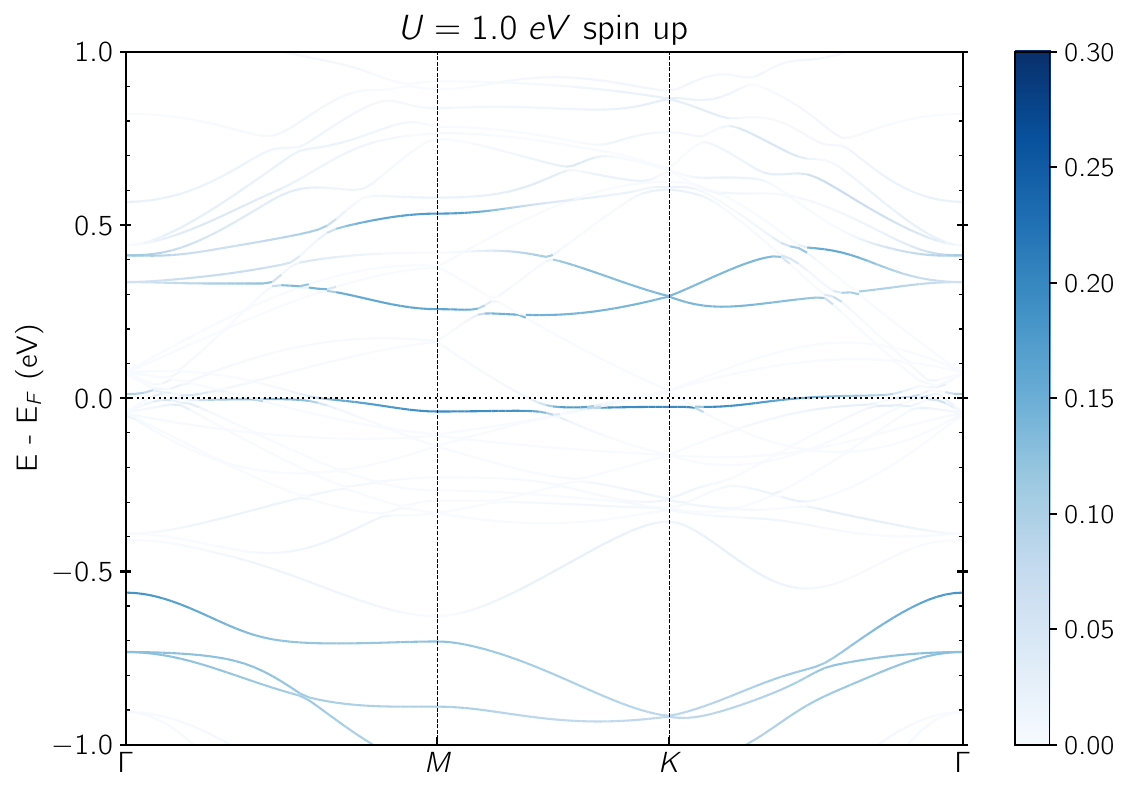}
\includegraphics[width=0.48\textwidth]{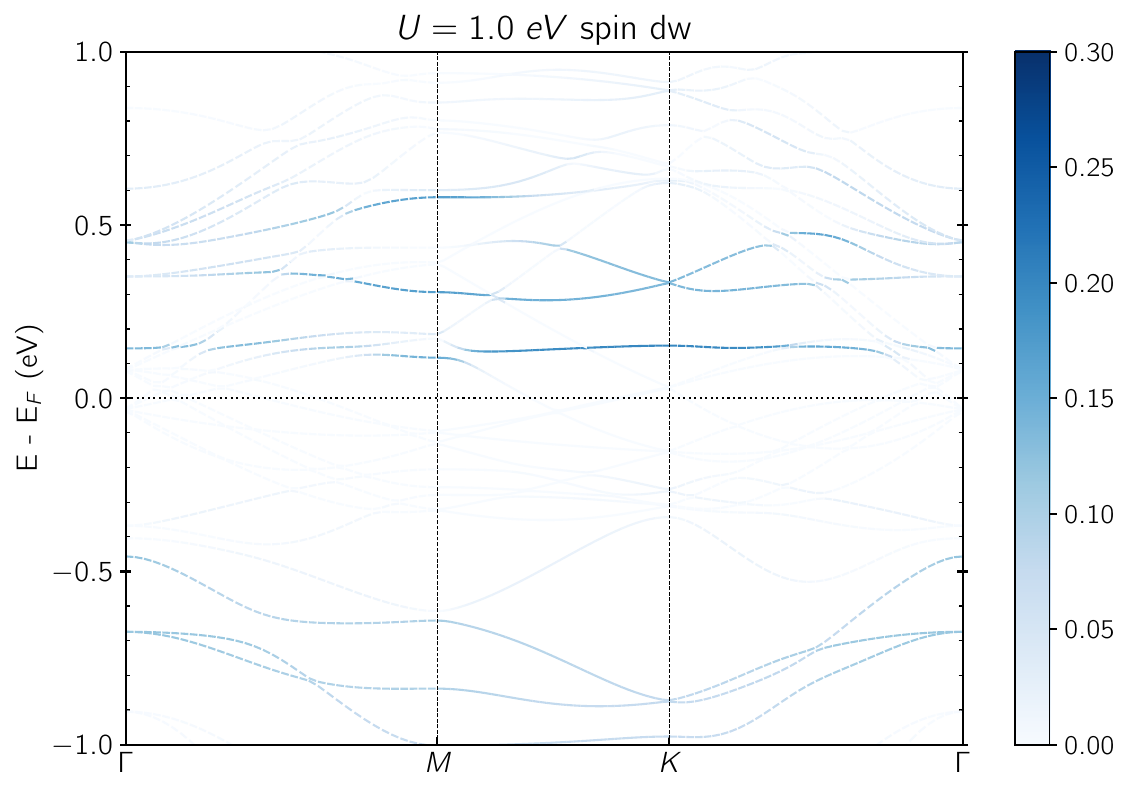}
\includegraphics[width=0.48\textwidth]{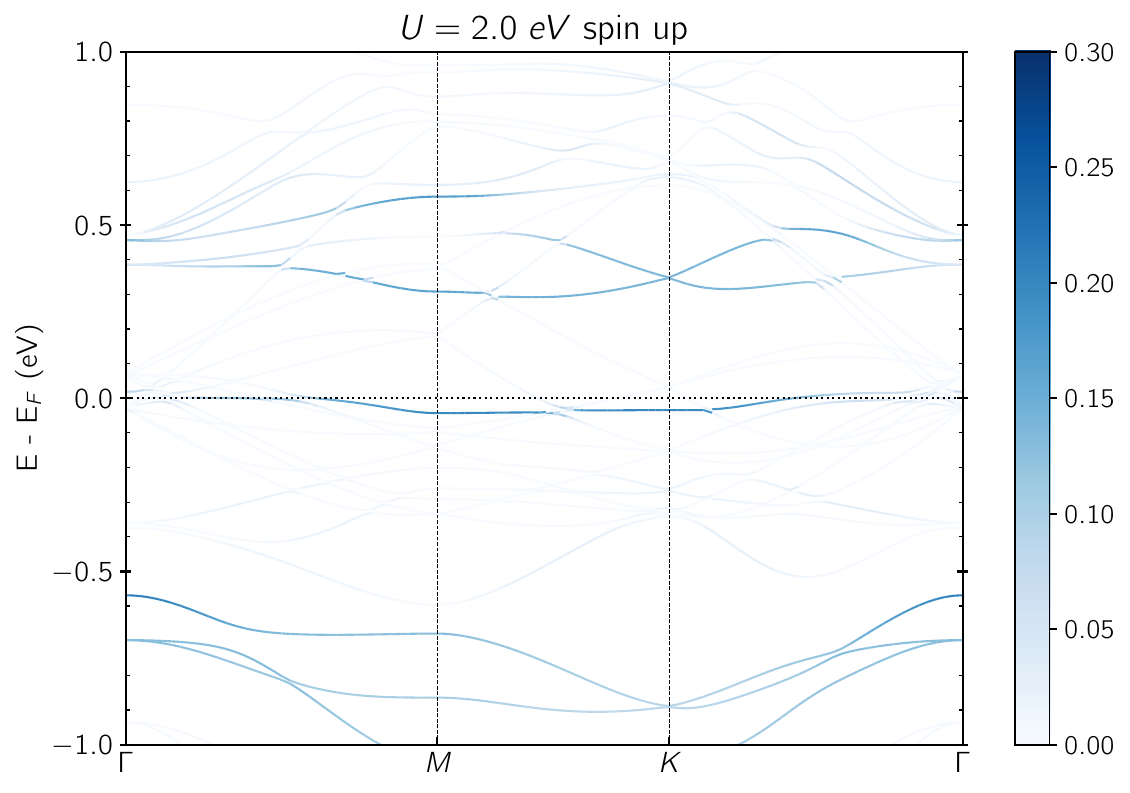}
\includegraphics[width=0.48\textwidth]{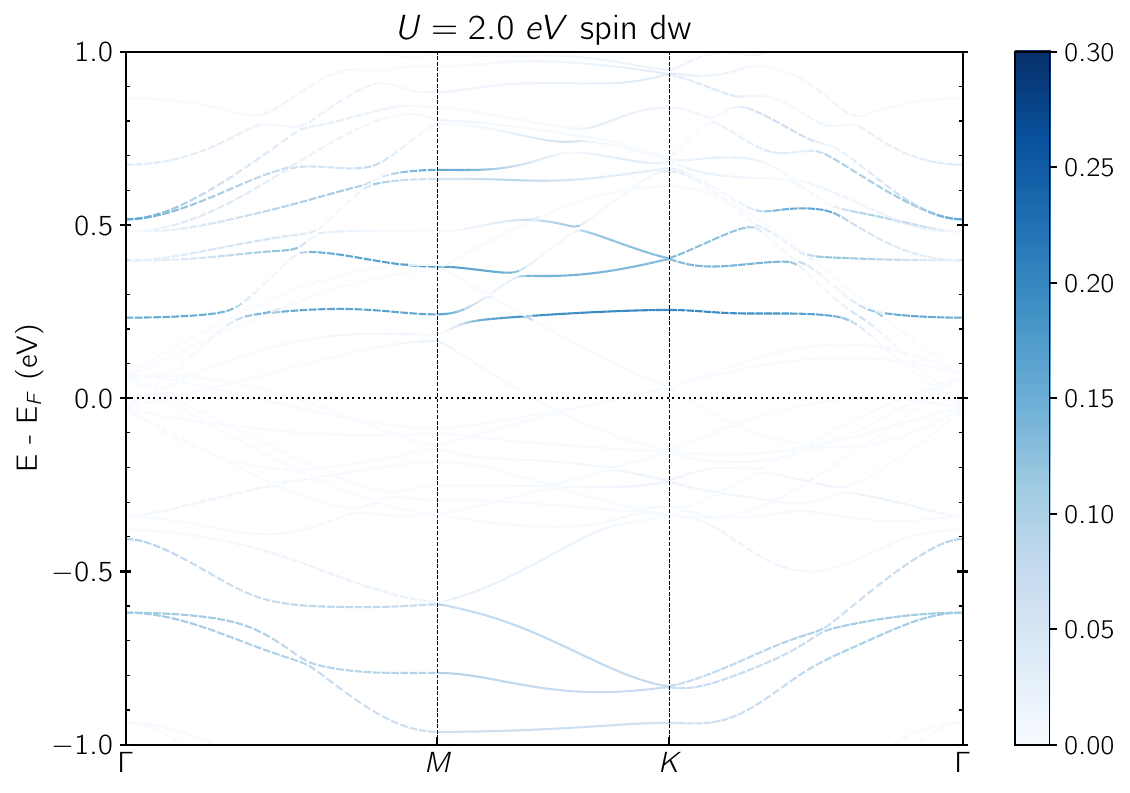}
\includegraphics[width=0.48\textwidth]{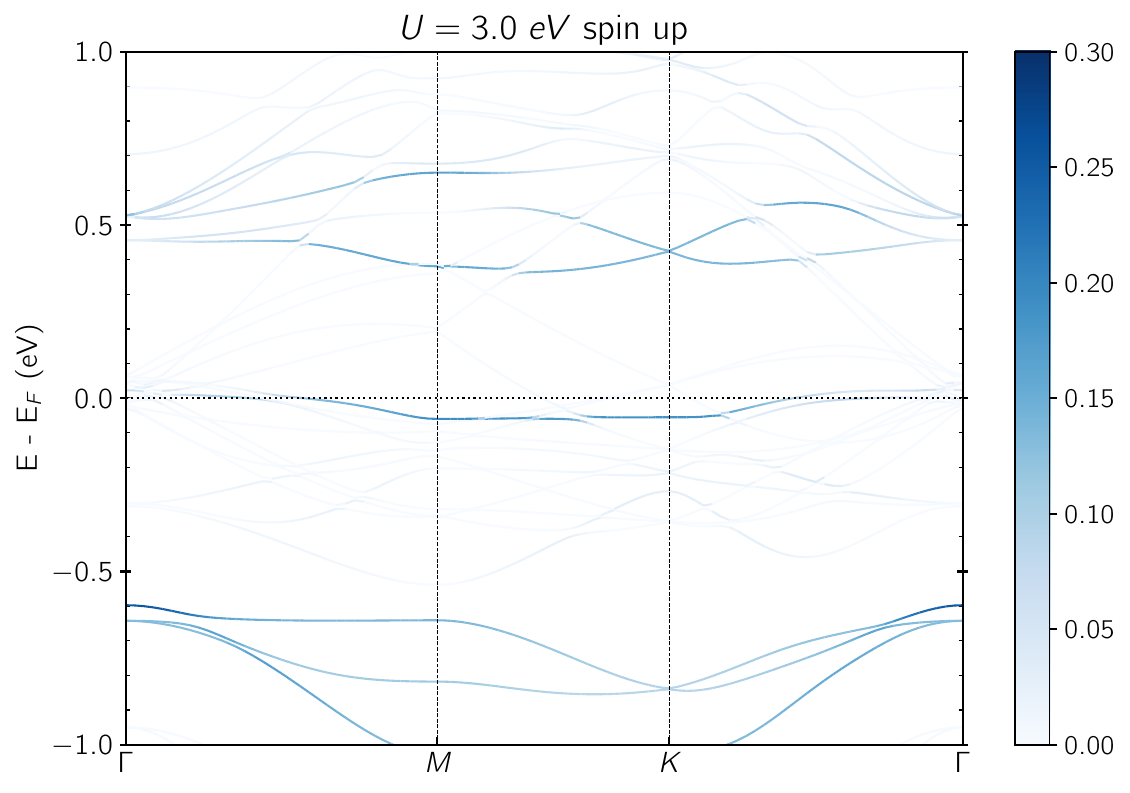}
\includegraphics[width=0.48\textwidth]{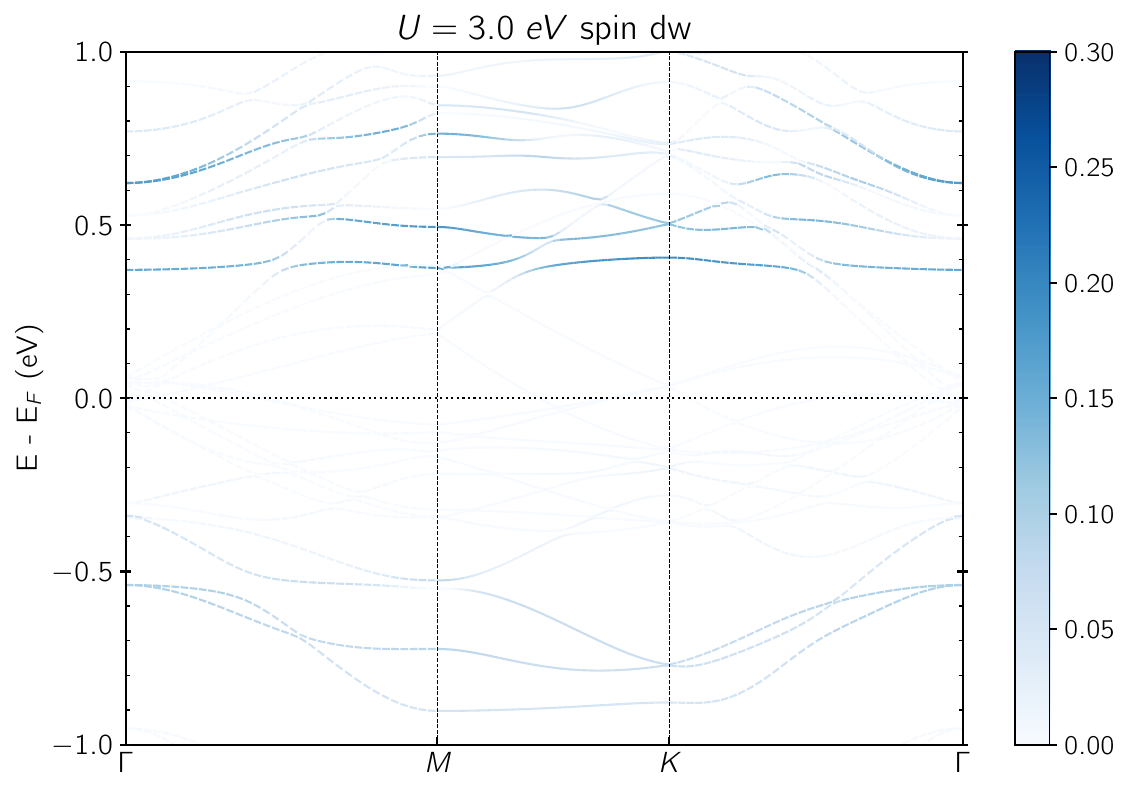}
     \caption{T-H TaS$_2$ electronic atom-projected (T A atom) bandstructures for the different values for $U$ considered in the main text and both spin polarizations.}
    \label{UDependence}
\end{figure*}

In this section, we include the spin-projected electronic bandstructures from TaS$_2$ bilayer for all four $U$ different values, see Fig.~\ref{UDependence}. We can determine that the overall effect of U is to split the T flat band, pushing the spin up band below the Fermi energy and the spin down one up into valence bands, thus increasing magnetization of this band and a hindering of the charge transfer.

%\begin{figure*}[t]
%    \centering
%\includegraphics[width=0.48\textwidth]{4hbtase2.png}
%\includegraphics[width=0.48\textwidth]{4hbtase2.png}
%     \caption{\noteFJ{Mock Fig.} Band structures of the 4H$_b$ compounds when the 1T layer is in the CDW state. The SoD centers for both T layers are on top of each other. Projected on Ta T vs H layers.}
%    \label{4H$_b$}
%\end{figure*}

\bibliography{TMD}

\end{document}